\documentclass[twocolumn]{aastex62}

\usepackage{mathrsfs}

\newcommand{\beq}{\begin{equation}}
\newcommand{\seq}{\end{equation}}

\newcommand{\gv}[1]{\ensuremath{\mbox{\boldmath$ #1 $}}} 

\graphicspath{{./}{figures/}}

\usepackage[symbol*]{footmisc}
\DefineFNsymbolsTM{myfnsymbols}{
  \textasteriskcentered *
  \textdagger    \dagger
  \textdaggerdbl \ddagger
  \textsection   \mathsection
  \textbardbl    \|%
  \textparagraph \mathparagraph
}%
\setfnsymbol{myfnsymbols}
   
\submitjournal{ApJL}

\shorttitle{A physical basis for LOC modeling}
\shortauthors{Waters \& Li}

\begin{document}

\title{The AGN broad line region as a clumpy turbulent outflow:\\ a physical basis for LOC modeling}

\correspondingauthor{Tim Waters}
\email{tim.waters@unlv.edu}

\author{Tim Waters}
\affil{Department of Physics \& Astronomy, University of Nevada, Las Vegas}
\author{Hui Li}
\affiliation{Theoretical Division, Los Alamos National Laboratory}

\begin{abstract}
Many studies have considered the roles of clouds, outflows, and turbulence in producing the broad emission and absorption lines in the spectra of active galactic nuclei (AGNs).  However, these are often treated as separate or even competing models.  Here, we consider the possibility that AGN clouds are condensations formed within the thermally unstable zones of outflows and then compare the typical sizes of such condensations with the injection scale $k_0^{-1} \sim L_0$ of turbulence, where $L_0$ is assumed to be the scale height of a representative global outflow model.  We find that for broad line region (BLR) parameters, clouds are many orders of magnitude smaller than $L_0$ and this has the following implication: BLR cloud dynamics can be modeled using a local approximation through the use of multiphase turbulence simulations of X-ray irradiated plasmas.
We present the first such 3D local clumpy turbulent outflow simulations.  We show that the condensations share the same type of
selection effects characterizing the locally optimally emitting cloud (LOC) scenario, thereby
offering a physical interpretation for the LOC model and accounting for its almost uncanny successes. 
The ubiquitous presence of emission line regions in AGNs can be simply explained as the natural outcome of there being a multiphase interval of $k$-space within the inertial range of a turbulent cascade.

\end{abstract}


\section{Introduction} 
The idea that AGN disk winds may be clumpy due to less ionized gas condensing, via thermal instability (TI), out of the highly ionized plasma comprising the bulk of the outflow dates back at least four decades (Davidson \& Netzer 1979).  
The first dynamical clumpy wind models by Krolik \& Vrtilek (1984) and Shlosman, Vitello, \& Shaviv (1985) were built upon pioneering theoretical studies of two-phase models in AGNs (Krolik, McKee, \& Tarter 1981) and Compton heated winds (Begelman, McKee, \& Shields 1983).  
Several authors further emphasized the role played by turbulence in these winds (e.g., Shields, Ferland, \& Peterson 1995; Chelouche \& Netzer 2005).   
In recent years, a clumpy turbulent outflow (CTO) scenario has increasingly been used to interpret observations of  
obscuring outflows (Kaastra 2014; Mehdipour et al. 2017; Turner et al. 2018), 
ultrafast outflows (e.g., Kraemer et al. 2018; Reeves et al. 2018), 
and quasar broad absorption lines (e.g., Krongold et al. 2017; Hamann et al. 2019; Leighly et al. 2019).  

In addition to these frequent invocations of a CTO picture, the 
overall importance of disk winds for explaining the diversity of AGN is well recognized (e.g., Giustini \& Proga 2019), 
and efforts have been made to show that the disk wind framework can be successfully used for reverberation mapping of the broad line regions (BLRs) of AGNs (e.g., Waters et al. 2016; Mangham et al. 2017).
Nevertheless,
there is no general acceptance of models suggesting that the BLR is due to condensations produced within disk winds (e.g., Czerny 2019).  
On the contrary, 
the most common view of the BLR is arguably 
one that arose out of the need to construct dynamical models to compare with reverberation mapping observations, namely the
notion that a population of pressure confined clouds orbits the central engine on quasi-Keplerian orbits (e.g., Netzer 2008).  
Such discrete orbiting cloud models are still regularly utilized for observational modeling (e.g., Grier et al. 2017; Gravity Collaboration et al. 2018; Murchikova et al. 2019) despite it having long been appreciated that any such clouds will be accelerated radially outward by the powerful radiation forces in AGN and transferred into the hot phase, thereby forming a wind (e.g., Mathews \& Ferland 1987). 

A common framework for modeling the BLR using photoionization calculations invokes the locally optimally emitting cloud (LOC) picture introduced by Baldwin et al. (1995), which by design is not tied to any particular physical scenario (see Ferland 2003).  
LOC models have been remarkably successful in reproducing emission line strengths and line ratios across a variety of prominent lines in both the BLR (e.g., Korista \& Goad 2000) and the narrow line region (NLR; see e.g., Ferguson et al. 1997) and are by now routinely used to model AGN variability (e.g., Guo et al. 2019; Korista \& Goad 2019).
In this Letter, we demonstrate that a CTO model of the BLR will give rise to the same type of selection effects that underly the success of LOC models.  

We proceed by first establishing in \S{2} that CTOs applied to the BLR can be modeled using local rather than global simulations.
In \S{3} we present our multiphase turbulence simulations designed to capture BLR cloud dynamics.
In \S{4} we show that the density statistics of these simulations are fully consistent with the LOC picture.  
We conclude by mentioning the agreement with other observational constraints in \S{5}.

\begin{figure*}
\includegraphics[width=\textwidth]{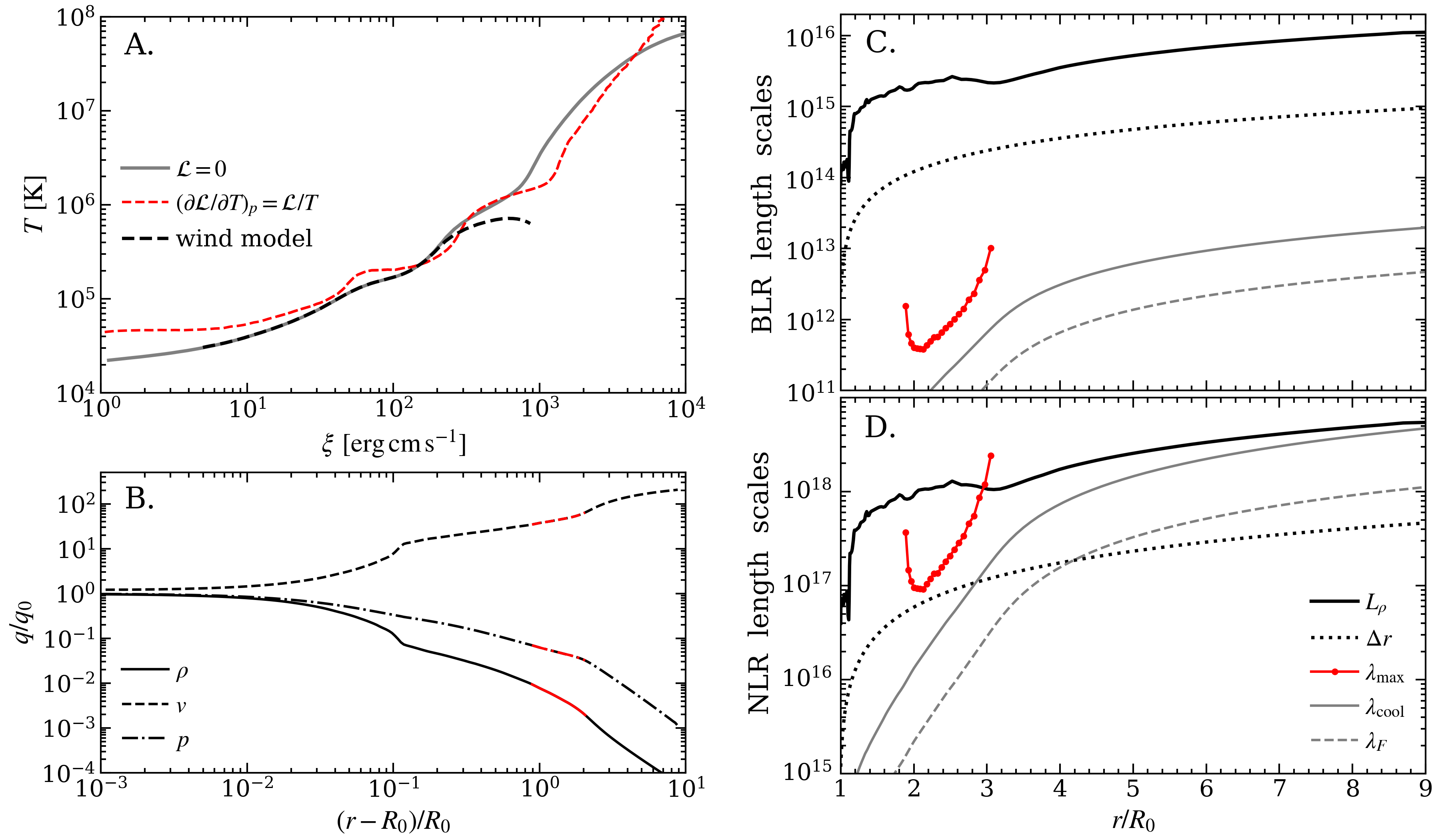}
\caption{
Analysis of an outflow solution containing a thermally unstable zone.    
Panel~A:  Phase space plot showing our representative AGN outflow solution (dotted black line), which follows the S-curve (gray line) corresponding to the AGN1 SED of Mehdipour et al. (2015).  The region of this solution that passes above the Balbus contour (dashed red line) is thermally unstable.  
Panel~B: Radial profiles of the outflow solution; thermally unstable zones are shown in red.
Panel~C: Comparison of the density scale height of the outflow solution scaled to the BLR (solid black line) with the grid scale (dotted line), the wavelength of the fastest growing TI mode (red line), the cooling length (solid gray line), and the Field length $\lambda_F$ (dashed gray line; the cloud evaporation scale - see Begelman \& McKee 1990).
Panel~D: Same as panel~C but for NLR parameters.  Since $\lambda_{\rm{max}} \ll L_\rho$ in the BLR, a local modeling approach is justified.} 
\end{figure*} 

\section{Cloud formation in a BLR outflow}  
While there have been several studies showing that TI leads to multiphase accretion flows (Barai et al. 2012; Gaspari et al. 2012; Mo{\'s}cibrodzka \& Proga 2013),
it has yet to have been demonstrated using hydrodynamical simulations that outflow solutions can be clumpy due to TI (although see Dannen \& Proga, in preparation).
To assess if this is simply a matter of unresolving the characteristic size scale of the clumps, 
we begin by analyzing an outflow solution that is formally thermally unstable. 
Namely, we compute scale lengths for a steady state spherically symmetric AGN outflow solution that features a thermally unstable zone on a phase diagram of temperature vs. ionization parameter (the $[T,\xi]$-plane hereafter, where $\xi = L_X/n r^2$ is the ionization parameter, with $L_X$ the ionizing luminosity, $n$ the hydrogen number density and $r$ the radius).  This solution, provided to us by R. Dannen \& D. Proga, is similar to the solutions published by Dyda et al. (2017) except (i) it is both a thermally and line-driven wind, as it includes the radiation force due to spectral line opacity that must accompany the heating due to irradiation (see Dannen et al. 2019); and (ii) it assumes irradiation falling off as $1/r^2$, whereas Dyda et al. (2017) used an isotropic radiation field.  The thermal driving is computed self-consistently from the unobscured NGC 5548 SED derived by Mehdipour et al. (2015) using photoionization calculations (see Dyda et al. 2017).  The associated radiative equilibrium curve, hereafter denoted the S-curve, which is the contour $\mathcal{L}=0$ (where $\mathcal{L}$ is the net cooling rate derived from the photoionization calculations), is shown as the solid black line in panel~A of Fig.~1.

The above outflow solution is plotted in Fig.~1 both on the $[T,\xi]$-plane (panel~A) and as a function of normalized radius, $x \equiv (r - R_0)/R_0$ (panel~B), where $R_0$ is the location of the base of the wind.  The minimum values of $(\xi,T) \equiv (\xi_0, T_0)$ correspond to $r=R_0$; the wind becomes hotter and more ionized as it expands.  
Notice that the solution stays on the S-curve until it undergoes significant expansion at $r \gtrsim 3 R_0$, beyond which it lies underneath the S-curve in a region of net heating, as required to balance the adiabatic cooling.  
The dashed red line is the `Balbus contour', i.e. the contour where $(\partial\mathcal{L}/\partial T)_p = \mathcal{L}/T$.
As shown by Balbus (1986), the local instability criterion for TI first found by Field (1965), $(\partial\mathcal{L}/\partial T)_p < 0$, only holds for points on the S-curve, whereas Balbus' criterion $(\partial\mathcal{L}/\partial T)_p < \mathcal{L}/T $, also holds for points off the S-curve.  
This instability criterion is formally satisfied for any points lying \emph{above} the Balbus contour on the $[T,\xi]$-plane, corresponding to the red portion of the profiles in panel~B.  For outflow solutions, this is a necessary but not a sufficient condition for TI; the flow dynamics can still stabilize TI (see Balbus \& Soker 1989). 
The question we want to address is whether or not the points in the thermally unstable zone of this solution can undergo local TI if the numerical resolution was orders of magnitude higher.  

The answer to this question depends critically on the size of condensations produced from TI relative to $L_q \equiv |q/\nabla q|$, the scale lengths characterizing the global outflow solution, where $q$ denotes any of the global spatial wind profiles $\rho(\gv{x}),\gv{v}(\gv{x}),p(\gv{x})$.
From Fig.~1, the density has the steepest gradient in the red region, so we will focus on $L_\rho$. 
In panel~C of Fig.~1, we compare $L_\rho$ (solid black curve) with the fastest growing wavelength of TI, $\lambda_{\rm{max}}$ (red curve), which provides an approximate upper limit to actual cloud sizes.  The value of $\lambda_{\rm{max}}$ is a function of density along the wind profile and is obtained numerically by solving the dispersion relation for TI (see e.g., Waters \& Proga 2019) 
assuming a wind solution with $R_0 = 1\,\rm{ld} \approx 2.6\times 10^{15}\,\rm{cm}$ and $n_0 \equiv n(R_0) =  10^{11}\,\rm{cm^{-3}}$, parameters typical of the inner radius of the BLR.
The comparison provides a clear answer to the above question: clumps are indeed sub-grid physics for this outflow simulation since $\lambda_{\rm{max}} \ll  \Delta r$ (the grid spacing).

For parameters characteristic of the NLR, on the other hand, we obtain $\Delta r \lesssim \lambda_{\rm{max}} \lesssim L_\rho$ (see panel~D of Fig.~1).  
The reason for this difference is easily understood.  The radiative equilibrium curve on the $[T,\xi]$-plane permits widely separated regions to share similar ionization states, but the scale heights of the global wind profiles depend on the actual distances to the emission regions.  For a given luminosity and ionization parameter, the base density varies as $n_0 \propto R_0^{-2}$, meaning that $L_\rho$ scales as $(d\ln n_0/dR_0)^{-1} \propto n_0^{-1/2}$.  Meanwhile, the characteristic cloud sizes scale with the cooling length, $\lambda_{\rm{cool}} \equiv c_s\,t_{\rm{cool}}$ (with $t_{\rm{cool}}$ defined in Fig.~2), 
which for a plasma with $\gamma = 5/3$ and solar abundances evaluates to
\beq \lambda_{\rm{cool}} \approx 3.3\times 10^{10}\,T_5^{3/2}\, n_9^{-1}\, \mathscr{L}_{23}^{-1}\: \rm{cm} ,\seq
where $T_5 = T/10^5\,\rm{K}$, $n_9 = n/10^9\,\rm{cm^{-3}}$, and $\mathscr{L}_{23}$ is the cooling rate in units of $10^{-23}\,\rm{erg\,cm^3\,s^{-1}}$.  
Thus, we have $\lambda_{\rm{cool}}/L_\rho \propto n_0^{-1/2}$, i.e. there is an increasingly large scale separation at large densities.

%
%

\section{Multiphase Turbulence Simulations}
The above calculations justify making an enormously helpful simplification, circumventing the need to perform global modeling to understand BLR cloud dynamics.  \emph{In a CTO model of the BLR, the local TI approximation holds since} $\lambda_{\rm{max}} \ll L_\rho$.   Thus, we can model the clumpy wind dynamics by applying a  standard tool --- isotropic turbulence simulations --- to assess the idea that a multiphase turbulent cascade can account for the properties of BLRs.     
Self-consistency requires that the size of the computational domain (`box-size' or $L_{\rm{box}}$ hereafter) satisfy $L_{\rm{box}} \ll L_\rho$ while simultaneously being larger than $\lambda_{\rm{max}}$ (thereby ensuring that doubling the box-size will not appreciably change the dynamics).

Whereas standard compressible turbulence simulations solve the equations of adiabatic hydrodynamics, we solve the equations of \emph{non-adiabatic} hydrodynamics, i.e. we include the physics of TI --- heating and cooling (H/C) processes and thermal conduction (TC), thus permitting the possibility that condensations can both form (via $\mathcal{L}$, the net H/C function) and evaporate (through TC; the evaporation scale is plotted as $\lambda_F$ in Fig.~1). 
Using such local CTO simulations, we now show that BLR clouds as we conceive of them form via TI only over a small range of wavenumbers in the inertial range of a turbulent cascade.

For our simulations, we use the `Blondin' S-curve (Blondin 1994) that contains the same H/C processes as the one computed for NGC 5548 by Dannen et al. (2019) but is analytic instead of tabulated and has been well tested numerically by Proga \& Waters (2015; PW15 hereafter).  Using the Athena code (Stone et al. 2008), we adopt the same fiducial physical parameters as PW15, only now we run 3D simulations for various box sizes (PW15 runs were 2D with a box size $L_{\rm{box}} = \lambda_{\rm{cool}}$).  Also, instead of the radiation force in the momentum equation of PW15, we apply a standard turbulence forcing prescription to drive the cascade process, assumed to have been initiated from scales larger than $L_{\rm{box}}$; see the review by Brandenburg \& Nordlund (2011) for the details of this modeling framework.  
We consider purely solenoidal forcing using a driving routine developed by Cho \& Lazarian (2002).
The strength of the turbulence is controlled by the turbulent Mach number $M_t = \overline{\delta v}/c_s$ (ratio of the RMS velocity amplitude due to forcing and the background adiabatic sound speed).  Broad emission lines show little evidence for shock heating (e.g., Ferland et al. 1996), so we consider a range of subsonic values $M_t = 0.05 - 0.75$.  
Our fiducial box size (runs A and B in Fig.~2) is $L_{\rm{box}} = 4\, \lambda_{\rm{cool}}$, large enough to include the fastest growing modes of TI ($\lambda_{\rm{max}} \approx 2\, \lambda_{\rm{cool}}$ for our initial values $\xi_0 = 1.9\times10^2 \,\rm{erg\,cm\,s^{-1}}$ and $T_0 = 1.9\times10^5\,\rm{K}$).  Our runs have a fixed resolution of $\Delta x = 0.02\, \lambda_F$.
 The Field length is $\lambda_F = 0.19 \,\lambda_{\rm{cool}}$ (see PW15), and we apply periodic boundary conditions.

\begin{figure}
\includegraphics[width=0.48\textwidth]{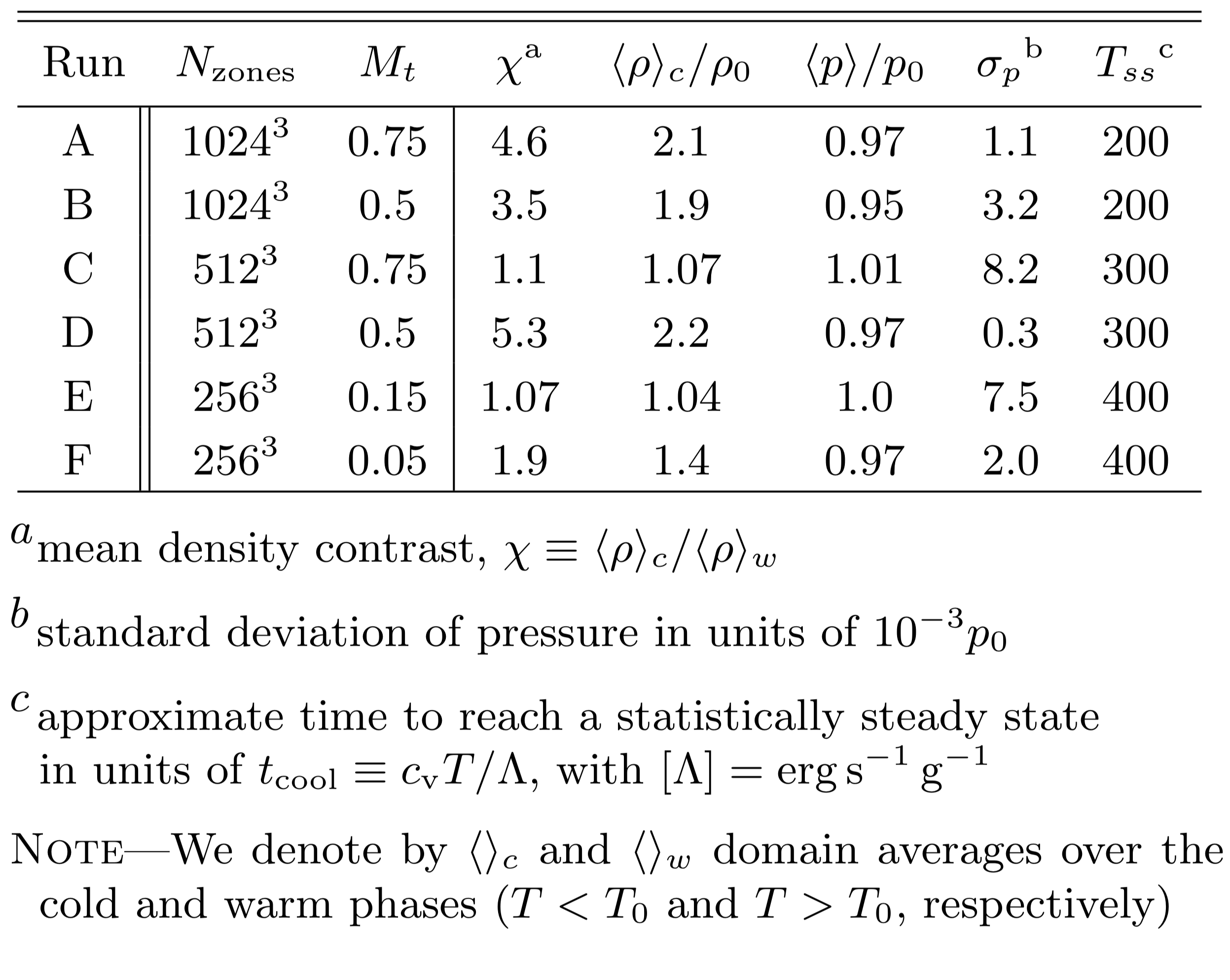}
\includegraphics[width=0.48\textwidth]{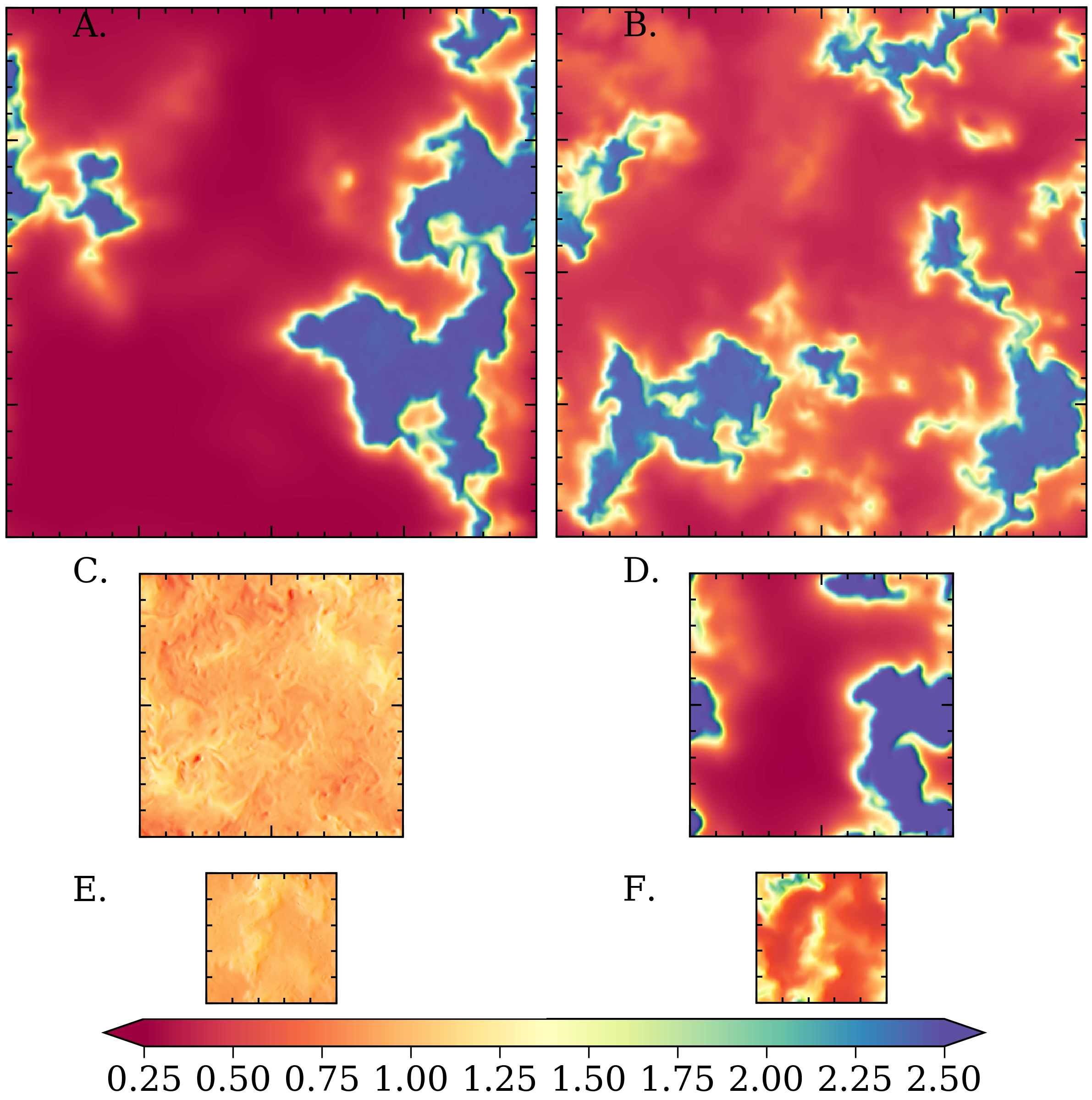}
\caption{
Summary/visualization of our multiphase turbulence runs demonstrating that cloud formation occurs within the inertial range of a turbulent cascade.
Panels~A-F show 2D slices of the density in units of $\rho_0$.   
Runs~(A/B,C/D,E/F) have box sizes $L_{\rm{box}} = (4,2,1)\, \lambda_{\rm{cool}}$, respectively. 
} 
\end{figure}

Fig.~2 presents the results of six different runs.  They altogether show that cloud formation is limited to the narrow range of wavenumbers $k_{\rm{min}} \leq k \leq k_F$, despite the gas being thermally unstable to all wavelengths exceeding $\lambda_F = 2\pi/k_F$.  Ideally, we would like to identify $k_{\rm{min}}$ by brute force by running ever larger box sizes.  Due to the difficulty of such an analysis and the computational expense of these simulations, we instead take advantage of the fact that the dispersion relation of TI has a maximum at $\lambda_{\rm{max}}$, meaning there is a value of $k$ in the range $2\pi/\lambda_{\rm{max}} \leq k \leq k_F$ with a growth rate $n_R$ satisfying $n_R(k) = n_R(k_{\rm{min}})$; see, e.g., Fig.~1 of Waters \& Proga (2019).  We therefore ran simulations with progressively smaller box sizes --- runs C-F in Fig.~2.  Run C has a box size $2\rm{x}$ smaller, revealing that a multiphase medium cannot develop for $M_t = 0.75$.  Only if we reduce $M_t$ can the slower growing TI modes in these smaller boxes condensate (compare runs D-F).  However, by the nature of a turbulent cascade, the velocity fluctuations are stronger at smaller $k$ (in incompressible turbulence, $\delta v \propto k^{-1/3}$); the energy injection rate is constant per unit volume and therefore 
$M_t$ would be increased not reduced for $k_{\rm{min}}$.  
By this reasoning, cloud formation in multiphase turbulence begins at some $k_{\rm{min}} \sim 2\pi/\lambda_{\rm{max}}$ because larger boxes will naturally suppress the growth of 
slow TI modes, i.e. those with $\lambda \gg \lambda_{\rm{max}}$.  

\subsection{Competing processes and turbulence statistics}
It is interesting to note that if these turbulence simulations could have been performed half a century ago after the discovery of TI, 
one could have predicted both the existence and the dynamics of BLR clouds.  However, we expect that our simulations are only capturing the leading order dynamics, as they neglect radiation and magnetic forces.  Radiation forces are known to be important in multiphase gas where resonance line opacity can increase by orders of magnitude as the condensations first appear (PW15).  This effectively provides a source of local kinetic energy injection, whereas the assumption in a turbulent cascade is that the dominant energy injection occured at much larger scales ($\sim L_\rho$ in Fig.~1).  The H/C term is also a source of (thermal) energy injection, and more work is needed to understand its relevance.  Grete et al. (2019) has already explored the effects of H/C in simulations not showing a multiphase transition.  Their simulations show that while turbulent dissipation can be balanced by the cooling source term, the kinetic, thermal, and magnetic energy spectra are quite insenstitive to the thermodynamics.  However, it has been argued that the non-barotropic nature of H/C processes prevent inverse cascades in 2D simulations (Hennebelle, \& Audit 2007).  We plan to address such issues in a followup study presenting the results of multiphase magnetohydrodynamic (MHD) simulations.

\section{Discussion}
In \S{2} we showed that BLR clouds are actually sub-grid physics for the resolutions obtainable in global simulations.
In \S{3} we demonstrated that BLR dynamics can be captured using local CTO simulations, greatly simplifying modeling efforts.
Here we discuss how the dynamics of the multiphase turbulent cascade is consistent with an LOC model. 
Bottorff \& Ferland (2002) considered a similar notion, but the BLR clouds were assumed to occupy the dissipation range rather than the inertial range and it was likely cost prohibitive to perform 3D multiphase turbulence simulations at the time.  

Underlying a typical LOC model is a large grid of photoionization calculations spanning many orders of magnitude in both ionizing flux (a proxy for distance) and hydrogen number density (Korista et al. 1997; see Leighly \& Casebeer 2007 for a review).  Within this wide 2D parameter space, the commonly observed emission lines span a relatively narrow range of ionization parameters; rather than $\xi$, most past studies use the dimensionless parameter
$U \equiv (\Phi_H/c)/n$, where $\Phi_H/c$ is the number density of hydrogen ionizing photons. 
For the NGC 5548 SED used in our Fig.~1 calculations, the conversion (determined using XSTAR) is $U \approx \xi/42$. 
If the clumps arise in optically thin plasma, 
the peak emissivities of prominent optical/UV emission lines typically fall in the range $-1.5 \leq \log(U) \leq 1$ (Shields et al. 1995).  
In ionization bounded clouds, $\log(U)$ can be smaller but rarely falls below $-3$.  

\begin{figure}
\includegraphics[width=0.46\textwidth]{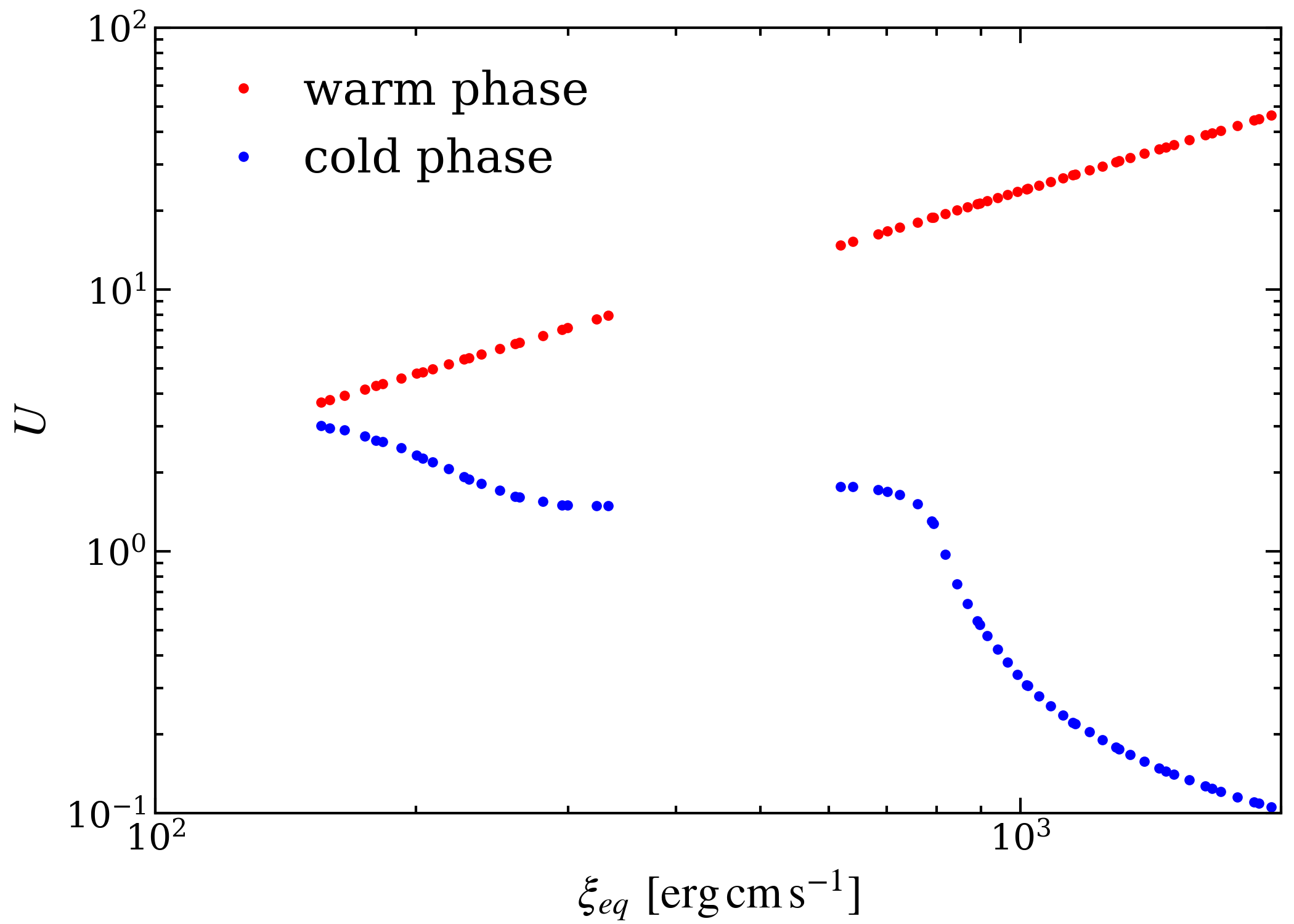}
\caption{Mapping from the unstable warm branch to the stable cold branch for the S-curve corresponding to the SED of NGC 5548.  
The red dots denote points within the two TI zones shown in panel~A of Fig.~1, while the blue dots mark the stable cold phase points connected to the unstable points by isobaric paths.  A local CTO simulation determines the distribution of gas at a given $\xi_{eq}$ and will span the entire vertical range connecting any pair of red and blue dots (with a small spread in $\xi_{eq}$).
This property implies that CTO solutions for the BLR are a realization of an LOC model.} 
\end{figure}

In Fig.~3, we plot the values of $U$ corresponding to the stable `cold branch' and unstable `warm branch' of TI on our S-curve in Fig.~1, i.e. the low-ionization points on the S-curve (blue dots) connected by isobaric paths ($45^\circ$ lines in panel~A) to the thermally unstable zones (red dots).  The gap in the data with $250 \leq \xi_{eq} \leq 500$ corresponds to the stable region in Fig.~1 where the Balbus contour rises above the S-curve thereby creating upper and lower TI zones.  Our representative outflow solution only occupies the lower TI zone (the left set of points in Fig.~3), but this was merely an example solution used to establish the validity of the local TI approximation.
Provided future global models of CTOs will show that the upper TI zone can be populated, it is valid to initialize local models at any of the red points in Fig.~3.  Local CTO simulations will then show that the entire range of $U$ between a given red point and its vertically connected blue point will be populated --- this defines gas in the cold and intermediate (or evaporating) phases.  

There will also be even more highly ionized gas defining the hot phase (see PW15 and Waters \& Proga 2019).  In this sense it is possible to obtain a large range of densities, temperatures, and ionization parameters from what began as a narrow range of outflow parameters. Thus, local CTO models provide a physical interpretation for the LOC scenario

\section{Concluding Remarks}
This work has shown that the BLR can be stratified according to some global wind solution and yet the local clump dynamics should be unaffected by the background wind gradients due to a large scale separation between gradient scale heights and cloud sizes.  The `global' picture accompanying our local CTO simulations is automatically consistent with ionization stratification in the BLR.  That is, reverberation mapping observations of NGC 5548 have established that $U$ must decrease radially outward (e.g., Peterson 1993), and the blue points in Fig.~3 show precisely this (since $\xi_{eq}$ increases with radius).  
Local multiphase turbulence simulations further reveal how a constant supply of line-emitting gas can be maintained in an environment hostile to a long lived population of BLR clouds: a statistically steady-state balance can be struck between cloud formation and evaporation.  Moreover, one of the main constraints on BLR cloud models --- the seemingly enormous number of clouds ($N_{\rm{cl}} \gtrsim 10^7$ or $10^8$) required to produce smooth line profiles (e.g., Arav et al. 1997) --- is easily satisfied.  Assuming typical cloud sizes are $\lambda_{\rm{cool}}$, panel~C in Fig.~1 shows that $10^{11}\,\rm{cm} \lesssim \lambda_{\rm{cool}} \lesssim 10^{12} \,\rm{cm}$ in the TI zone,  
so for $R_0 = 10^{16}\,\rm{cm}$ and $N_{\rm{cl}} \sim (C_f R_0/\lambda_{\rm{cool}} )^3$, we have for a covering fraction $C_f = 0.1$, $10^{9} \lesssim N_{\rm{cl}} \lesssim 10^{12}$.

\bigskip 
\acknowledgments
We thank Daniel Proga for his comments on an early draft of the manuscript.
We additionally thank him and his graduate student Randall Dannen for providing us with the representative
outflow solution used in this study and for making the results of their XSTAR photoionization calculations
for the NGC 5548 SED of Mehdipour et al. (2015) publicly available.  We are grateful to Shengtai Li for sharing his turbulence forcing routine.  TW further acknowledges fruitful discussions with Kirk Korista on LOC models and Philipp Grete on performing non-adiabadic turbulence simulations.
H.L. acknowledges the support of the LANL/LDRD program, the NASA/ATP program, and the DoE/OFES program.

\end{document}